\documentclass[aps,prl,twocolumn,longbibliography]{revtex4-1}
\usepackage{bm}
\usepackage{color} 
\usepackage{graphicx}
\usepackage{epsfig}
\usepackage{hyperref}
\usepackage{natbib}
\usepackage{amsmath}
\usepackage{amssymb}
\usepackage{bbm}
\newcommand{\Sec}[1]{\section{#1}}

\newcommand{\rmi}{{\rm i}}

\newcommand {\e}{{\rm e}}

\begin{document}

\title{Persistent spin grids with spin-orbit coupled 2D electron gas}

\author{A.~V.~Poshakinskiy}
\email[E-mail:~]{poshakinskiy@gmail.com}

\affiliation{ICFO-Institut de Ciencies Fotoniques, The Barcelona Institute\\ of Science and Technology, 08860 Castelldefels, Barcelona, Spain}

\begin{abstract}

We consider the diffusive spin dynamics of a 2D electron gas with spin-orbit coupling confined within a grid of narrow channels. We show that the lifetime of certain spin distributions in such grids greatly exceeds that in an unconfined 2D electron gas and diverges as the channel width approaches zero. Such persistent spin grids occur if the electron spin orientation remains invariant after diffusion around the grid plaquette. We establish a topological $\mathbb{Z}_2$ classification for persistent spin grids and speculate that the setup can be used to simulate non-Abelian lattice gauge theories.

\end{abstract}

\maketitle

\Sec{Introduction}%
In semiconductor nanostructures, relaxation of the free electron spin polarization is predominantly governed by the Dyakonov--Perel mechanism~\cite{Dyakonov71}, involving spin precession in an effective spin-orbit magnetic field that typically depends linearly on the electron momentum~\cite{Dyakonov86}. This field arises inherently in nanostructures lacking spatial inversion symmetry and has two main contributions~\cite{Ganichev2014}: the Dresselhaus contribution from bulk crystal inversion asymmetry~\cite{Dresselhaus1955} and the Rashba contribution from structural inversion asymmetry~\cite{Bychkov84}. When electrons propagate diffusively between two points, they follow various trajectories, each with a different spin rotation angle. Averaging over these trajectories leads to spin relaxation.

Spin relaxation can be suppressed in two ways. First, by precisely matching Rashba and Dresselhaus parameters, the spin rotation angle becomes trajectory-independent, resulting in emergent SU(2) symmetry~\cite{Bernevig2006}. This symmetry gives rise to a spin density wave known as the persistent spin helix (PSH), which has a significantly longer lifetime compared to a homogeneous spin distribution~\cite{Weber2007,Koralek2009}. The helical spin pattern can also emerge naturally during spin diffusion~\cite{Gridnev2002}, e.g., after the initial excitation of spin polarization by a focused optical beam~\cite{Walser2012,Ishihara2014,Ishihara2022,Ishihara2023}, and can be manipulated using external electric and magnetic fields~\cite{Altmann2016,Kohda2017,Anghel2018,Passmann2019,Anghel2021}.

The second approach involves lateral confinement of a 2D electron gas. Spin lifetime is predicted to increase as $1/w^2$ when the channel width $w$~\cite{Kiselev2000,Malshukov2000} or dot size~\cite{Aleiner2001} becomes smaller than the spin-orbit precession length (but is larger than the electron mean free path). Experiments confirm this enhancement in narrow channels and wires, with limitations due to cubic Dresselhaus terms~\cite{Holleitner2007,Altmann2014,Altmann2015,Eberle2021}. The underlying physical mechanism is that small spin rotations stemming from motion across the channel commute and cancel, while rotations due to motion along the channel persist and give rise to a long-lived helical spin pattern independently of the PSH condition~\cite{Slipko2011,Wenk2011,Altmann2014,Altmann2015}.

In real structures, tuning to the PSH condition is often impractical, or can undesirably alter other parameters~\cite{Anghel2022}. While lateral dots and channels do not need the tuning, they preclude spin transport or restrict it to one dimension. Here, we propose a structure where 2D diffusive spin transport with suppressed spin relaxation can be achieved for spin-orbit parameters far detuned from the PSH condition.
We consider a 2D electron gas subjected to a lateral potential that constrains electron diffusion to a grid of channels. Narrow channels suppress spin relaxation within them, while a properly designed grid geometry ensures that electrons diffusing between two points experience the same spin rotation angle regardless of the path taken on the grid. As a result, a persistent spin grid is formed, enabling spin transport in arbitrary directions with minimal spin relaxation. When an electron diffuses around a plaquette of such a grid, its spin undergoes an integer number of full rotations. Depending on the parity of this number, spin grids can be classified into two topologically distinct classes.

\Sec{Spin diffusion in a grid}%
We consider 2D electron gas with spin-orbit interaction of the form $H_{\rm SOI} =  \bm h(\bm k) \cdot \bm \sigma$, where $\bm \sigma$ is the vector of Pauli matrices and  $\bm h(\bm k)$ depends linearly on the electron wave vector $\bm k$. For the (001)-oriented zinc-blend quantum wells (QWs), the latter has the form~\cite{Ganichev2014}  $\bm h (\bm k) = (\beta+\alpha)k_y \bm e_x + (\beta-\alpha)k_x \bm e_y $, where $\alpha$ and $\beta$ are the Rashba and Dresselhaus parameters respectively, $\bm e_x \parallel [1\bar{1}0]$ and $\bm e_y \parallel [110]$ are the in-plane directions, and $\bm e_z \parallel [001]$. 
If an electron propagates ballistically in the $x$($y$) direction, its spin rotates due to the effect of the spin-orbit interaction with the frequency $(2/\hbar) \bm h$, making a full rotation at the spin-orbit length $|\lambda_{x,y}|$, where $\lambda_{x,y} =\pi\hbar^2/[m^*(\beta \pm \alpha)]$ and $m^*$ is the electron effective mass.

However, in most real structures, electron exhibits multiple scattering events at the length of $|\lambda_{x,y}|$. 
Then, the evolution of the spin polarization density $\bm S(\bm r,t)$, $\bm r=(x,y)$, is described by a diffusion equation~\cite{Malshukov2000,Froltsov2001,Gridnev2002,Kleinert2007,Poshakinskiy2016cor,Ferreira2017,Schliemann2017,Passmann2019} that can be presented in the form
\begin{align}\label{eq:diff}
\frac{\partial \bm S(\bm r,t)}{\partial t }= D \sum_{\alpha=x,y}\left(  \frac{\partial}{\partial r_\alpha} - \bm \Lambda_{\alpha} \right)^2  \bm S(\bm r,t) \,,
\end{align}
where $D$ is the spin diffusion coefficient and $\bm\Lambda_{x,y}$ are 3$\times$3 skew-symmetric matrices  that describe the rotation of  moving spins in the spin-orbit field. In the general case, $\Lambda_\alpha^{il} = 2\sum_j \epsilon_{ijl} \langle h_j(\bm k)  k_\alpha  /E_k \rangle$, where the angular brackets denote the averaging over electron wave vectors $\bm k$ with an appropriate distribution, $E_k = \hbar^2 k^2/{2m^*}$. For (001)-oriented QWs,  
\begin{align}\label{eq:lambda}
\bm \Lambda_{x,y} \bm S=  2\pi\lambda_{x,y}^{-1}\, \bm e_{y,x} \times \bm S \,.
\end{align}
In Eq.~\eqref{eq:diff}, we neglect the additional spin relaxation channel due to $k$-cubic spin-orbit interaction, which is typically small and insensitive to lateral confinement~\cite{Malshukov2000,Altmann2014,Altmann2015}.

Note that the differential operator in the right-hand side of Eq.~\eqref{eq:diff} has a formal similarity with the Hamiltonian of a vector particle in a SO(3) gauge field $\bm \Lambda_\alpha$. Importantly, we describe here the semiclassical evolution of spin density vector~\cite{Malshukov2000}, in contrast to quantum approaches where the spin-orbit field is regarded as an SU(2) gauge field for the electron wave function~\cite{Frohlich1993,Hatano2007,Chen2008,Yang2008,Tokatly2008,Tokatly2010a}. 

\begin{figure}[t!]
    \centering
    \includegraphics[width=0.83\columnwidth]{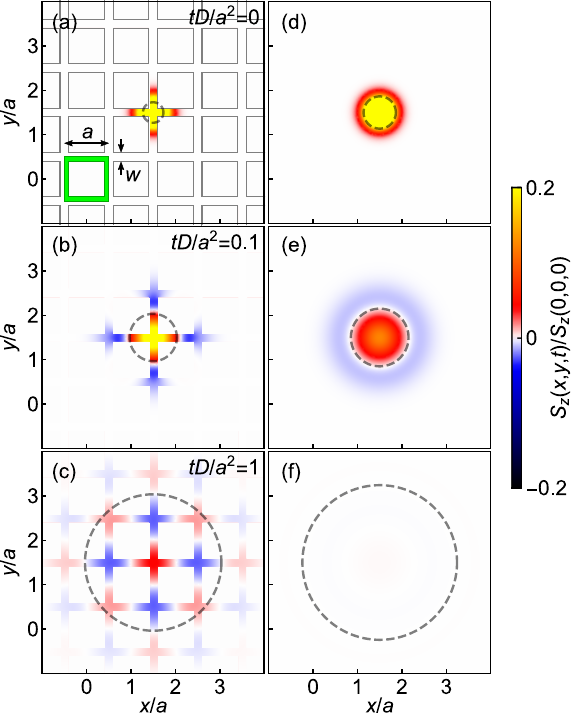}
    \caption{(a,b,c) Spin diffusion in a square grid as compared to (d,e,f) the unconfined spin diffusion. The panels show spatial distributions of $S_z$ at different times, indicated in the plot. Dashed circles indicate the spin distribution spread and have the radii $\big[\int r^2|\bm S(\bm r, t)| d^2 r/ \int |\bm S(\bm r, t)| d^2 r\big]^{1/2}$. Pure Rashba or Dresselhaus spin-orbit interaction is assumed. The grid has the period $a= \lambda_{x,y}/2$ and the edge width $w/a=0.2$. Unit cell of the grid is shown by green color in panel (a). Initial distribution of $S_z$ at $t=0$ is assumed Gaussian with the standard deviation $\sigma=0.2a$.
    }
    \label{fig:time}
\end{figure}

We suppose that the external lateral potential restricts electron motion to a grid-like region $G$, e.g., the square grid with the period $a$ and the edge width $w$ depicted in Fig.~\ref{fig:time}(a). If no additional spin relaxation occurs at the boundaries of the region $\partial G$,  the spin current across the boundary should vanish, leading to the boundary condition~\cite{Malshukov2000}
\begin{align}\label{eq:bc}
\sum_{\alpha=x,y}n_\alpha \left( \frac{\partial}{\partial r_\alpha} - \bm \Lambda_{\alpha}\right) \bm S(\bm r,t) \Big|_{\partial G} = 0 \,,
\end{align}
where $n_\alpha$ is the normal to the boundary. 

Figure~\ref{fig:time} shows the evolution of the $S_z$ spin density after its excitation at $t=0$ in a small area in the center, obtained from the numerical solution of Eqs.~\eqref{eq:diff},\eqref{eq:bc} for the case of pure Rashba or pure Dresselhaus spin-orbit field, $\lambda_{x} = \lambda_{y}$. Panels (a,b,c) show the spin diffusion on a square grid with the period $a= \lambda_{x,y}/2 $ which corresponds to a half of a full spin rotation in the spin-orbit field. Panels (d,e,f) show as a reference the spin diffusion of an unconfined gas in the same color scale. In both cases, spatial spin-density oscillations emerge due to the spin-orbit interaction. In the case of unconfined diffusion, one can barely see a single oscillation before the spin polarization vanishes. In contrast, the spin polarization in the grid survives for much longer times, revealing many spatial oscillations, cf. Figs.~\ref{fig:time}(c) and (f). Note that the spatial spread of the spin distributions (dashed lines) is similar in the two cases~\footnote{The effective diffusion coefficient on the square grid can be estimated as $D_{\rm eff} = Da/(2a-w)$.}.
The suppression of spin relaxation in the grid, while keeping the 2D character of the diffusion, is the main result of this work. In what follows, we define the conditions under which such a long-lived spin grid emerges and quantify its lifetime.

%%%%%%%%%%%%%%%%%%%%%%%%%%%%%%%%%%%
%%%%%%%%%%%%%%%%%%%%%%%%%%%%%%%%%%%

%%%%%%%%%%%%%%%%%%%%%%%%%%%%%%%%%%%
\begin{figure*}[t!]
    \centering
    \includegraphics[width=0.99\textwidth]{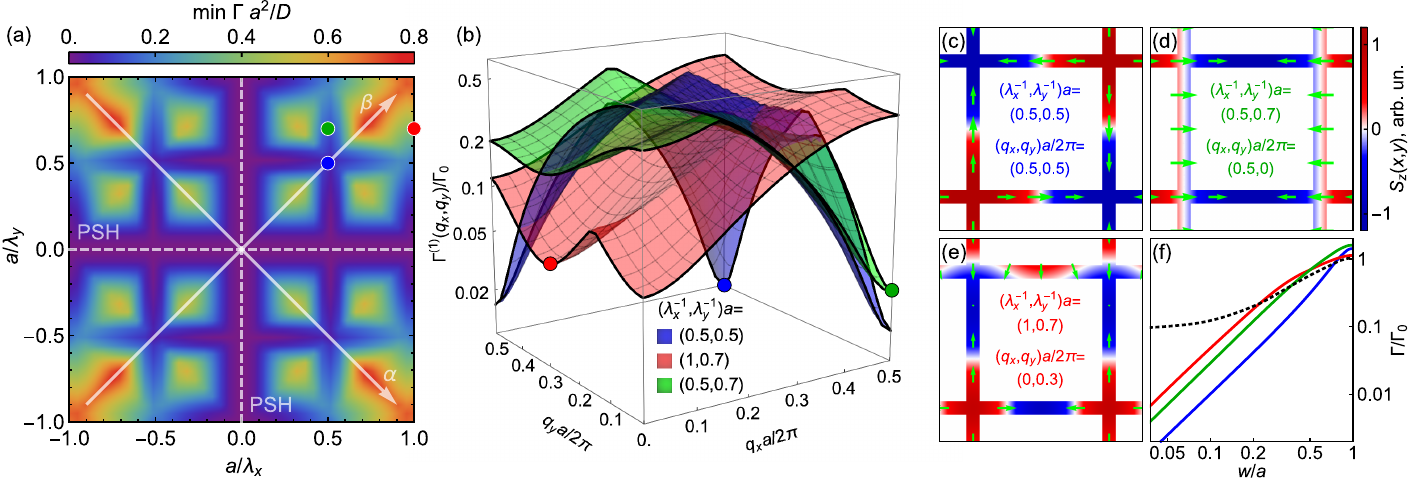}
    \caption{(a) The decay rate of the most long-lived spin-diffusion Bloch mode in a square grid with the period $a$ as a function of spin-orbit parameters $\lambda^{-1}_x$, $\lambda^{-1}_y$. Dashed lines indicate the PSH conditions for unconfined 2D electron gas. (b) Dispersion of the lowest decay band for three different configurations, which are labeled in the plot and marked by points in panel (a). (c)--(e) Spin distributions for the long-lived spin-diffusion Bloch modes marked by points in panels (b). The color encodes $S_z(x,y)$, while $[S_x(x,y)$, $S_y(x,y)]$ are shown by arrows. (f) Scaling of the Bloch-mode decay rates with the grid edge width $w$. Solid blue, green, and red curves correspond to the long-lived modes marked by the dots of the same color in panel (b). Black dashed curve is for the mode with $(q_x , q_y) /2\pi = (\lambda^{-1}_x , \lambda^{-1}_y)  = (0.7,0.7)/a$. 
    For panels (a)--(e), the grid edge width $w=0.1a$ was chosen. In panels (b) and (f), the decay rates are normalized by the decay rate of the most long-lived spin mode  in the unconfined electron gas, $\Gamma_{0} = D (q_</4)^2 [8 - (q_</q_>)^2]$, where $q_< = 2\pi \,{\rm min\,}(|\lambda^{-1}_x|, |\lambda^{-1}_y|)$, $q_> =  2\pi \,{\rm max\,}(|\lambda^{-1}_x|, |\lambda^{-1}_y|)$~\cite{Passmann2019}. 
    }
    \label{fig:disp}
\end{figure*}
%%%%%%%%%%%%%%%%%%%%%%%%%%%%%%%%%%%

\Sec{Spin-diffusion Bloch modes}%
 Since we study spin diffusion in a periodic grid, it is essential to consider the Bloch modes, i.e., the eigenfunctions of the right-hand side of Eq.~\eqref{eq:diff} in the unit cell of the square grid, $ (a-w)/2 \leq |x|, |y| \leq a/2$, shown by green color in Fig.~\ref{fig:time}(a), with the boundary condition Eq.~\eqref{eq:bc} on the inner cell boundary and the boundary condition $\bm S(a/2,y) = \bm S(-a/2,y) \e^{\rmi q_x a}$,  $\bm S(x,a/2) = \bm S(x,-a/2) \e^{\rmi q_y a}$ on the outer cell boundary, where $x$ and $y$ are measured from the center of the grid plaquette and $q_x$, $q_y$ are quasi wave vectors. Note that the differential operator in the right-hand side of Eq.~\eqref{eq:diff}, accompanied by the aforesaid boundary conditions, is self-adjoint. Therefore, it has real eigenvalues, which form  the decay rates bands $\Gamma^{(n)}(q_x,q_y)$, similar to the electron energetic bands in a crystal.
 
 The decay of spin polarization at large times is determined by the minimal value of the lowest band, ${\rm min}\,\Gamma^{(1)}$. In Fig.~\ref{fig:disp}(a), we show how this minimal value depends on the spin-orbit parameters $\lambda_{x,y}^{-1}$ (or $\alpha$ and $\beta$) for a fixed grid period $a$. As expected, the decay rate vanishes in the PSH conditions, $\lambda_{x}^{-1}=0$ or $\lambda_{y}^{-1}=0$ ($\alpha = \pm \beta$), see dashed lines. In addition, the map reveals other low-decay configurations, occurring when $a/\lambda_{x}$ or $a/\lambda_{y}$ are integer or half-integer. Away from these conditions, the general trend is the increase of the decay with the increase of the spin-orbit parameters. 

Now we explore three particular low-decay configurations, indicated by colored dots in Fig.~\ref{fig:disp}(a),
and for them calculate the decay dispersion in the lowest band $\Gamma^{(1)}(q_x,q_y)$, see Fig.~\ref{fig:disp}(b). The blue surface corresponds to $(\lambda^{-1}_x , \lambda^{-1}_y) a =(0.5,0.5)$ [blue dot in  Fig.~\ref{fig:disp}(a)]. There, the minimal decay rate is achieved at $(q_x,q_y)/2\pi =  (\lambda^{-1}_x , \lambda^{-1}_y)$, see blue dot in  Fig.~\ref{fig:disp}(b). The corresponding mode has the spin distribution shown Fig.~\ref{fig:disp}(c), which matches the emerging long-lived pattern in Fig.~\ref{fig:time}. The dispersion also has two other slightly less deep minima at  $(q_x,q_y)/2\pi =  (0 , \lambda^{-1}_y)$ and $ (\lambda^{-1}_x ,0)$.

Now we detune the spin-orbit parameters to $(\lambda^{-1}_x , \lambda^{-1}_y)a =(0.5,0.7)$ [green dot in Fig.~\ref{fig:disp}(a) and green surface in Fig.~\ref{fig:disp}(b)]. Then, only the minimum at $(q_x,q_y)/2\pi = (\lambda^{-1}_x ,0)$ remains, see green dot in  Fig.~\ref{fig:disp}(b). The corresponding spin distribution has no spin oscillations along $y$ axis: when diffusing along $y$, the spin polarization appears to be parallel to the spin-orbit field, see Fig.~\ref{fig:disp}(d).  

We also consider the case $(\lambda^{-1}_x , \lambda^{-1}_y)a  =(1,0.7)$  [red dot in Fig.~\ref{fig:disp}(a) and red surface in Fig.~\ref{fig:disp}(b)]. Then, the minimal spin decay rate is achieved for the mode with $(q_x,q_y)/2\pi  \approx (\lambda^{-1}_x , \lambda^{-1}_y)$, which is equivalent to $(q_x , q_y) a/2\pi \approx (0,0.3)$, see red dot in  Fig.~\ref{fig:disp}(b). The corresponding mode can, in general, have the period that is incommensurate with the grid period, see Fig.~\ref{fig:disp}(e). 

Finally, we examine how the decay rate of the described modes scales with the grid edge width $w$, see blue, green and red curves in Fig.~\ref{fig:disp}(e), which correspond to the modes shown in Figs.~\ref{fig:disp}(c),~(d), and~(e), respectively. For all of them, we have $\Gamma \propto w^2$, similarly to the spin decay rate in a channel~\cite{Kiselev2000,Malshukov2000,Altmann2015}. In the limit $w \to 0$, the modes become decayless and form persistent spin grids. For comparison we also show by the black dashed line the decay rate of the spin mode with $(q_x , q_y) /2\pi = (\lambda^{-1}_x , \lambda^{-1}_y)  = (0.7,0.7)/a$, which has a finite decay rate in the limit $w \to 0$ and does not form a persistent spin grid.

%%%%%%%%%%%%%%%%%%%%%%%%%%%%%%%%%%%
\begin{figure}
    \centering
    \includegraphics[width=0.99\columnwidth]{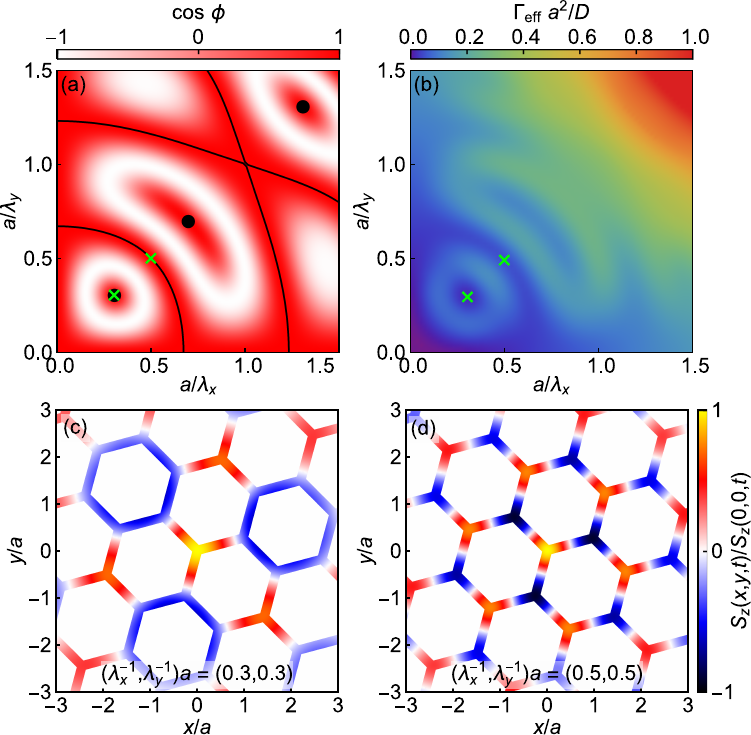}
    \caption{(a) Spin direction change upon the diffusion around a plaquette of the honeycomb lattice as a function of the spin-orbit-interaction parameters $\lambda_{x,y}^{-1}$. Black lines (points) show the solutions of $\cos \phi =1$, which correspond to (non-)trivial persistent spin grids. (b) The effective decay rate $\Gamma_{\rm eff}$ of the $S_z$ spin distribution as a function of $\lambda_{x,y}^{-1}$. 
  (c),(d) Spatial distribution of $S_z$ at time $t=5 a^2/D$ after the initial focused excitation in the center of the plot. Spin-orbit parameters are indicated in the plot and marked by green crosses in panels~(a),(b). The angle between an edge of the hexagon and a coordinate axis is $\pi/12$; the grid edge width is (b) $w=0.05a$, (c),(d) $w=0.2a$.}
    \label{fig:hex}
\end{figure}
%%%%%%%%%%%%%%%%%%%%%%%%%%%%%%%%%%%

\Sec{Classification of persistent spin~grids}%
If the grid edge width $w$ is much smaller than $|\lambda_{x,y}|$, the edges can be regarded as quasi-1D and the spin relaxation inside them is greatly suppressed~\cite{Malshukov2000,Altmann2015}. On the larger scale, when the electrons travel between two given points, they can still follow different paths on the grid. If the paths result in different final spin orientations, the averaging over them leads to spin relaxation. Therefore, for the persistent spin grid to occur, all paths on the grid should yield the same spin rotation.

This condition can be formalized using the matrix, similar to the Wilson loop in lattice gauge theories,
\begin{align}\label{eq:R}
\bm R = \mathcal{P} \exp \oint  \sum_{\alpha=x,y} \bm\Lambda_\alpha dr_\alpha \,,
\end{align}
where the integral is taken around a plaquette of the grid and $\mathcal P$ denotes the path ordering. Matrix $\bm R$ describes the rotation of  the spin polarization vector upon diffusion around the plaquette. As for any rotation, there always exists a spin direction, parallel to the rotation axis, that remains unchanged upon diffusion around the plaquette. However, for a persistent spin grid, one needs the spin direction to remain unchanged after diffusion around any number of plaquettes. In generic case, insetting an additional path around another plaquette into some point in the integral Eq.~\eqref{eq:R} changes the preserved spin direction, spoiling the persistent spin grid formation~\footnote{
In some specific cases, the persistent spin grid can be formed despite $\bm R \neq \bm 1$, if there exists a spin direction that is preserved or inverted upon the spin transfer to all points of the grid that are equivalent to the chosen starting point. Such situation happens in the square grid if $ a/\lambda_x$ or $a/\lambda_y$ is half-integer. In generic grids, this requires fine tuning of several parameters. 
}. 
However, if $\bm R =\bm 1$ for all plaquettes, adjoining a plaquette to an arbitrary point of a path does not change the spin rotation operator for the path. Therefore, for any number of plaquettes, $\bm R = \bm 1$, enabling formation of the persistent spin grid. From the lattice-gauge-theory perspective, it means that the gauge field $\bm\Lambda_\alpha$ can be eliminated by proper gauge transform. 

Since $\bm R$ is an element of the 3-dimensional SO(3) group, in general, tuning it to identity requires 3 adjustable parameters (such as the spin-orbit constants or the grid edge lengths). However, this task is simplified if the grid plaquette possesses certain symmetries. For instance, if the plaquette has an inversion center, one can split the path around it into two halves, $\bm R = \bm R_2 \bm R_1$, where $\bm R_2$ can be obtained from $\bm R_1$ by inverting the gauge field $\bm\Lambda_\alpha$, which leads to $\bm R_2 = \bm P \bm R_1 \bm P$, where $\bm P$ is the rotation by $\pi$ around the direction $\tilde z$ defined by $[\bm\Lambda_x,\bm\Lambda_y]$; in case of the spin-orbit interaction Eq.~\eqref{eq:lambda}, $\tilde z = z$. Then, for $\bm R = \bm 1$ it is sufficient that the rotation axis of $\bm R_1$ is perpendicular to $\tilde z$. The latter condition defines a 2-dimensional manifold inside SO(3), therefore, one can expect to fulfill it by tuning a single system parameter. 

As an example, we consider spin diffusion on a honeycomb grid. To find the spin-orbit parameters that enable persistent spin grids, we evaluate the invariant
\begin{align}\label{eq:cosf}
\cos \phi = \frac{{\rm Tr\, }\bm R -1}{2} \,,
\end{align}
where $\phi$ is the angle of spin rotation and $\cos \phi =1$ corresponds to $\bm R=\bm 1$.
Figure~\ref{fig:hex}(a) shows the color map of $\cos \phi$ as a function of the spin-orbit parameters $\lambda_{x,y}^{-1}$ for a fixed grid edge length $a$. The dependence is even in $\lambda_{x,y}^{-1}$ and quasiperiodic, i.e., is a sum of several incommensurate periodic functions corresponding to spin rotation in different edges of the plaquette. The black lines and dots show the persistent spin grid conditions, $\cos \phi =1$. 

In the persistent spin grid conditions, $\bm R=\bm 1$, Eq.~\eqref{eq:R} defines a loop path in the SO(3) group. Such loops, and the corresponding persistent spin grids, can be classified according to the fundamental group $\pi_1[\text{SO(3)}]=\mathbb{Z}_2$. Trivial loops can be contracted to a point and correspond to an even number of full rotations; non-trivial loops cannot be transformed to the trivial ones and correspond to an odd number of full rotations. Whether the loop is trivial or not can be determined by replacing in Eq.~\eqref{eq:R} the elements of the $\mathfrak{so}(3)$ algebra $\bm\Lambda_\alpha$ by their counterparts from $\mathfrak{su}(2)$, 
$\bm\sigma \cdot \bm\Lambda_\alpha \bm\sigma/4$. Then, the integration gives $\pm \bm 1$  for the (non-)trivial loops. 

The two types of the persistent spin grids are clearly identified in the map of $\cos \phi$, Fig.~\ref{fig:hex}(a). Those corresponding to trivial loops are located on black lines and are connected to the point $(\lambda_{x}^{-1},\lambda_{y}^{-1})=(0,0)$, which suggests the precise way to transform the loop into a point. Nontrivial persistent spin grids correspond to isolated black points. Since the trivial persistent spin grids are achieved on continues sets of spin orbit parameters, the corresponding spin patterns need not to be commensurate with the grid period and are more robust against changes of the spin-orbit parameters.

To test the predicted persistent spin grids, we calculate the evolution of the spin distribution $S_z(x,y,t)$ after the initial local excitation of $S_z$ at $x,y=0$. Then, we extract the effective decay rate from $\Gamma_{\rm eff} t = -\ln S_z(0,0,t)/S_z^{(0)}(0,0,t)$, where $S_z^{(0)}(x,y,t)$ stands for the pure spin diffusion in the absence of spin-orbit interaction and $t=5a^2/D$ was chosen. We show the map of $\Gamma_{\rm eff}$ as a function of spin-orbit parameters $\lambda_{x,y}^{-1}$ in Fig.~\ref{fig:hex}(b). The map follows the pattern of $\cos \phi$, cf. Fig.~\ref{fig:hex}(b) and~(a): The minima of $\Gamma_{\rm eff}$ correspond to $\cos \phi =1$, where the persistent spin grids are predicted. Note that with the increase of $\lambda_{x,y}^{-1}$, the minima of $\Gamma_{\rm eff}$ are smeared, because the spin relaxation within the grid edges grows rapidly as $(\lambda_{x}\lambda_{y})^{-2}$~\cite{Kiselev2000,Malshukov2000,Altmann2015,Grobecker2024arxiv}. Therefore, more favorable are the configurations with small $a/\lambda_{x,y}$, such as those marked by green crosses in Figs.~\ref{fig:hex}(a) and~(b). The corresponding long-lived spin patterns are shown in Figs.~\ref{fig:hex}(c),(d).

\Sec{Outlook}%
%
%
%%%%%%%%%%%%%%%%%%%%%
We have demonstrated that spin relaxation in a 2D electron gas can be suppressed by confining it to a properly designed grid of narrow channels. Such grids can be fabricated via chemical etching of quantum well structures~\cite{Altmann2014,Altmann2015} or by applying voltage to patterned gates. Tuning to the persistent spin grid conditions can be accomplished using gate voltages that control spin-orbit parameters~\cite{Dettwiler2017,Luengo2017}. To enhance spin grid excitation efficiency, an oscillating spin pattern around a plaquette can be generated with a single shot of a vector vortex beam~\cite{Ishihara2023} or programmable structured light~\cite{Kikuchi2024arxiv}.

The extended spin lifetime in the grid enables not only longer spin diffusion lengths but also to drift spins further in desired directions using in-plane electric fields. Investigating the effects of external electric and magnetic fields on persistent spin grids presents a promising direction for future research. Another avenue is the study of weak (anti)localization~\cite{Wenk2011,Saito2022} and quantum corrections to spin transport, where a distinct behavior might arise for the two topological classes of persistent spin grids, differing in the sign of the electron wave function after a trip around a plaquette.

From a fundamental perspective, spin dynamics in grids---where only the most long-lived mode survives at large times---could serve as a simulator of lattice gauge theories, performing imaginary time evolution to find ground states. In our analysis, the gauge field was fixed, determined by spin-orbit constants, while only the matter field evolved. At high excitation powers, however, spin-orbit parameters may become electron-density-dependent~\cite{Passmann2018}, enabling simulations of nonlinear models where both the gauge and matter fields are dynamic.

\Sec{Acknowledgments}%
%
%
%%%%%%%%%%%%%%%%%%%%%
We thank S. Anghel for fruitful discussions.

\end{document}